\documentclass[12pt]{article}
\usepackage{psfig}

\def\gtrsim{\mathrel{\mathpalette\vereq>}}
\makeatletter
\def\vereq#1#2{\lower3pt\vbox{\baselineskip1.5pt \lineskip1.5pt
\ialign{$\m@th#1\hfill##\hfil$\crcr#2\crcr\sim\crcr}}}
\makeatother

\begin{document}

\begin{titlepage}
\begin{center}
\today     \hfill    LBNL-42429 \\
~{} \hfill UCB-PTH-98/53  \\
~{} \hfill hep-ph/9810468\\

\vskip .1in

{\large \bf Tau Appearance In\\Atmospheric Neutrino Interactions}%
\footnote{This work was supported in part by the U.S. 
Department of Energy under Contracts DE-AC03-76SF00098, in part by the 
National Science Foundation under grant PHY-95-14797.  HM was also 
supported by Alfred P. Sloan Foundation.}

\vskip 0.3in

Lawrence J. Hall and Hitoshi Murayama

\vskip 0.1in

{\em Department of Physics\\
     University of California, Berkeley, California 94720}

\vskip 0.1in

{\em Theoretical Physics Group\\
     Ernest Orlando Lawrence Berkeley National Laboratory\\
     University of California, Berkeley, California 94720}

\vskip 0.05in

\end{center}

\vskip .1in

\begin{abstract}

  If the correct interpretation of the Super-Kamiokande atmospheric 
  neutrino data is $\nu_\mu \rightarrow \nu_\tau$ oscillation,
  the contained data sample should already have more than 10 $\tau$ 
  appearance events. We study the challenging task of detecting the
  $\tau$, focussing on the decay chain $\tau^\pm \rightarrow \rho^\pm
  \rightarrow \pi^\pm \pi^0$ in events with quasi-elastic $\tau$ production.
  The background level, which is currently quite
  uncertain because of a lack of relevant neutral current data,
  can be measured by the near detector in the K2K experiment.  Our 
  estimates of the background suggest that it may be possible to detect
  $\tau$ appearance in Super-Kamiokande with 5--10 years of running.
\end{abstract}

\end{titlepage}

\newpage

\section{Introduction}

Recent data from the Super-Kamiokande (SK) detector \cite{SK} has provided 
confirmation of the atmospheric neutrino anomaly, previously reported
by Kamiokande \cite{Kam} and IMB \cite{IMB} collaborations.  The Soudan-II
\cite{Sou} and MACRO \cite{MACRO} Collaborations have also reported
anomalies which 
support the Super-Kamiokande observations.  
Apparently about 40\% of the $\nu_\mu$ produced in cosmic ray showers
in the atmosphere do not reach the underground detectors.  While
calculations of the $\nu_\mu$ flux contains uncertainties, these
calculations can be checked by comparing with the observed rate of
muon and $\nu_e$ induced events. Neutrino oscillations provide a plausible
depletion mechanism which has been strengthened
by recent Super-Kamiokande data showing that the amount of 
$\nu_\mu$ depletion depends on the azimuthal angle.

In this letter, we argue that, if the oscillation is to $\nu_\tau$,
then the Super-Kamiokande detector could detect charged current $\tau$
appearance events via the chain $\nu_\tau \rightarrow \tau^-
\rightarrow \rho^- \rightarrow \pi^- \pi^0$. The signal event rate is
small because only a small fraction of $\nu_\tau$ have sufficient
energy to produce $\tau$. Furthermore, there is considerable
uncertainty in the background from $\rho$ production via other
neutrino induced processes. However, we estimate the signal to
background ratio to be above unity, and the signal has a strong
up-down asymmetry. A multi-year search at Super-Kamiokande with an 
efficiency of 50\% could lead to a $5 \sigma$ signal.

The atmospheric $\nu_\mu$ could oscillate into $\nu_e$,
$\nu_\tau$ or $\nu_s$, an SU(2) singlet neutral fermion.
Fits to the atmospheric data, in models with three generations and a single 
$\Delta m^2$ relevant to the atmospheric oscillation, show that at
most about a third of the oscillations can be to $\nu_e$, so that at
least two thirds are to $\nu_\tau$ \cite{barger}. For values of
$\Delta m^2$ above $2 \times 10^{-3}$ eV$^2$, the CHOOZ experiment
constrains the $\nu_e$ fraction to be even smaller
\cite{CHOOZ}. Within the minimal scenario, the oscillations are
dominantly to $\nu_\tau$. While oscillations to $\nu_s$ remain a
possibility, it is theoretically less favored.\footnote{It
requires the addition of an exotic fermion,
neutral under the known gauge interactions. Many theories contain
such fermions; however, because of their neutrality, they typically
get very large masses, thereby inducing small
masses for $\nu_{e,\mu,\tau}$ by the see-saw mechanism. On the other
hand, for $\nu_\mu$ to oscillate into $\nu_s$, the mass of $\nu_s$
must certainly be less than 250 keV. Understanding such a light mass
for a singlet state is possible, but requires more elaborate 
theoretical structures.}

How can the preferred $\nu_\mu \rightarrow \nu_\tau$ interpretation be 
verified? A long baseline experiment with low neutrino beam energy
({\it e.g.}\/, KEK to SK) would see $\nu_\mu$ disappearance but not 
$\tau$ appearance. For higher energy neutrino beams ({\it e.g.}\/, 
MINOS and ICARUS)
the more direct signal of $\tau$ appearance is possible. In this
letter we argue that a $\tau$ appearance signal, with a large up-down
asymmetry, may be possible at Super-Kamiokande. We caution that the
event rates are low, and that a significant $\rho$ detection efficiency
is necessary. It may be possible to see this signal even if $\Delta
m^2 \leq 10^{-3}$ eV$^2$, where the planned accelerator based
experiments become more difficult.

\vskip .25in

\section{$\tau$ Appearance}

In this section, we give numerical results for 
$\tau$-appearance cross sections, from both quasi-elastic (QE) processes
\begin{eqnarray}
         &  & \nu_{\tau} n \rightarrow \tau^{-} p,
        \label{eq:QEnu}  \\
         &  & \bar{\nu}_{\tau} p \rightarrow \tau^{+} n,
        \label{eq:QEnubar}
\end{eqnarray}
and deep inelastic scattering (DIS) processes
\begin{eqnarray}
         &  & \nu_{\tau} p \rightarrow \tau^{-} X ,
        \label{eq:DISnup}  \\
         &  & \nu_{\tau} n \rightarrow \tau^{-} X ,
        \label{eq:DISnun}  \\
         &  & \bar{\nu}_{\tau} p \rightarrow \tau^{+} X ,
        \label{eq:DISnubarp}  \\
         &  & \bar{\nu}_{\tau} n \rightarrow \tau^{+} X .
        \label{eq:DISnubarn}
\end{eqnarray}
We have not calculated the additional contributions from 
resonance production.  Therefore, the sum of the QE and DIS 
contributions gives a conservative estimate of the total 
$\tau$ production cross section.

\begin{figure}
  \centerline{\psfig{file=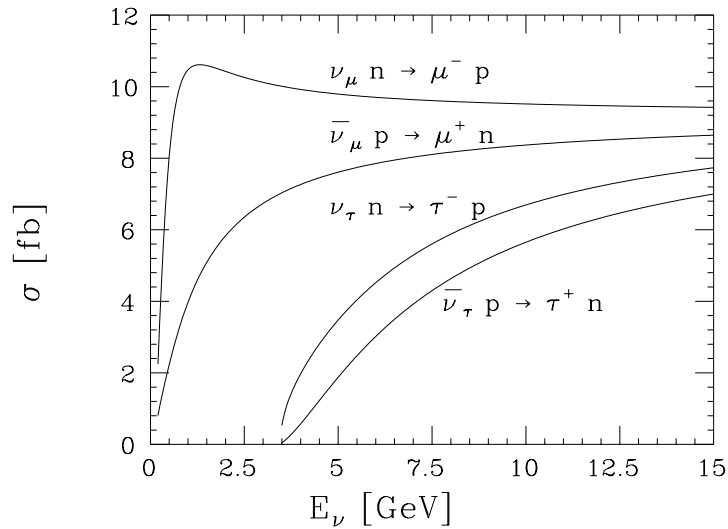,width=0.7\textwidth,angle=90}}
        \caption[cross]{The cross sections of $\tau$ production from 
          $\nu_{\tau}$ charged current reactions, for quasi-elastic
          (QE) events as described in the text.  The $\mu$ production
          cross sections are shown for comparison.  }
        \label{fig:cross}
\end{figure}

The QE cross sectionsare evaluated using the formula in 
Ref.~\cite{Llewellyn-Smith}, with the second-class currents dropped, 
PCAC and dipole form factors assumed, and taking $m_{A} \simeq 1.0$~GeV 
as suggested by experiments \cite{Kitagaki}.  The results are shown 
in Fig.~\ref{fig:cross} 
together with the corresponding quasi-elastic cross sections for 
muon (anti-)neutrino induced events.  We quote cross sections for 
$\tau$ production at $E_{\nu} = 5$~GeV:
\begin{eqnarray}
        \sigma_{QE} (\nu_{\tau} n \rightarrow \tau^{-} p)  & = &  
        3.5~\mbox{fb} ,
        \label{eq:QEtau-}  \\
        \sigma_{QE} (\bar{\nu}_{\tau} p \rightarrow \tau^{+} n) & = & 
        2.0~\mbox{fb} .
        \label{eq:QEtau+} 
\end{eqnarray}

\begin{figure}
  \centerline{\psfig{file=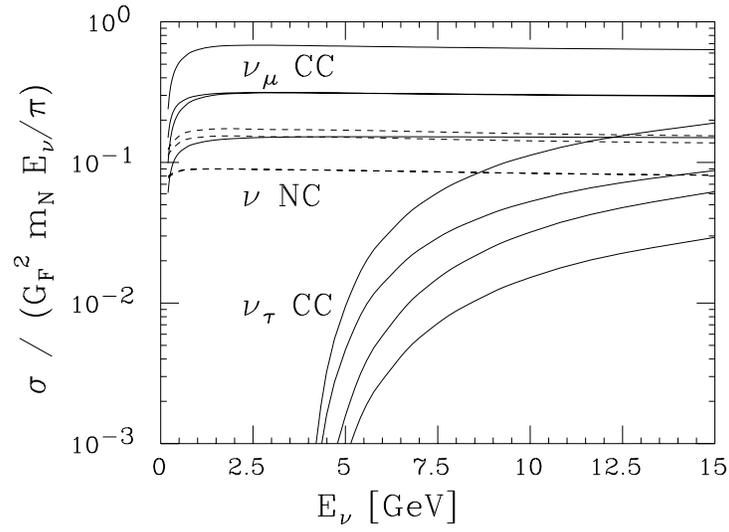,width=0.7\textwidth,angle=90}}
        \caption[cross]{The cross sections of $\tau$ production from 
          $\nu_{\tau}$ charged current reactions for deep-inelastic
          (DIS) events, normalized by $G_F^2 m_N E_\nu/\pi$.  The
          rising solid curves ($\nu_\tau$ CC) are for $\bar{\nu}_\tau
          n$, $\nu_\tau p$, $\bar{\nu}_\tau p$, and $\nu_\tau n$ from
          below.  The neutral current ($\nu$ NC) processes are given
          in dashed lines, for $\bar{\nu} p(n)$, $\nu p$, $\nu n$ from
          below; the first two are accidentally close and are not
          distinguishable in the plot.  The $\nu_\mu$ CC production
          cross sections are shown for comparison, for $\nu_\mu n$,
          $\nu_\mu (\bar{\nu}_\mu) p$, and $\bar{\nu}_\mu n$ from
          above, where the middle two are accidentally close and are
          not distinguishable in the plot.}
        \label{fig:DISratio}
\end{figure}

The DIS cross sections 
are evaluated using the MRS-A parton
luminosities, specifically extended to the low-$Q^{2}$ regime 
\cite{MRS-A}.  The parton-level cross
sections, however, are the leading-order calculations \cite{AJ}.
Cross sections for all four processes,
(\ref{eq:DISnup},\ref{eq:DISnun},\ref{eq:DISnubarp},\ref{eq:DISnubarn})
are shown in Fig.~\ref{fig:DISratio}.\footnote{In all of our discussions, we took the $\tau$
  threshold from nucleons at rest.  In reality, however, the Fermi
  motion of nucleons inside oxygen nuclei {\it lowers}\/ the threshold
  and hence the expected signal rate is higher.  We thank G. Shapiro
  on this point.  A precise estimate of this effect, however, requires
  a nuclear model and is beyond the scope of this paper.  It is worth
  emphasizing though that we estimate the signal rate conservatively
  only using QE process with fixed production threshold.}
  
The QE cross sections are combined with calculations of the 
atmospheric neutrino flux \cite{Honda} for a 414 day exposure at
SuperKamiokande (21.1~kton), assuming no nuclear effects.  The total number 
of electron events calculated is 106.7 for the multi-GeV sample, $p_{e} > 
1.33$~GeV. According to the MC calculations based on the same 
atmospheric neutrino flux \cite{Honda} shown in the 2nd paper in 
\cite{SK}, the
Super-Kamiokande collaboration expect 182.7 $e$-like events, with 38.3\% 
from the QE process.  Therefore, we overestimate the QE event rate,
due to our lack of knowledge of 
the detection efficiency and the nuclear shadowing effect, and we should
multiply our calculated results by a reduction factor of 0.66.

To confirm this reduction factor, we further compared 
the multi-GeV
$\mu$-like events to our calculations.  This is slightly more 
complicated because there are both fully contained (FC) and partially 
contained (PC) events.  We find that there should be 250.0 QE events.  
Their MC shows 229.0 FC events, with a 54.1\% QE fraction, and 287.7 PC events 
with 17.9\% QE: in total, 175.5 QE events.  The same reduction factor 
of 0.66 explains the descrepancy (actually about 0.7),
and hence we adopt this
factor to scale between our calculated rates and those from the
Super-Kamiokande Monte Carlo.

The Super-Kamiokande Collaboration observed 218 $e$-like multi-GeV events 
in 414 days. This is larger than the 182.7 events expected from their
Monte Carlo calculation by a factor of 1.19. We assume that this has
nothing to do with oscillations, but is due to the uncertainties in
the flux calculation. We therefore assume that all unoscillated
fluxes, produced in the cosmic ray showers, are a factor 1.19 larger
than calculated in \cite{Honda}, which increases our reduction factor
from 0.66 to 0.77.

\begin{figure}
  \centerline{ \psfig{file=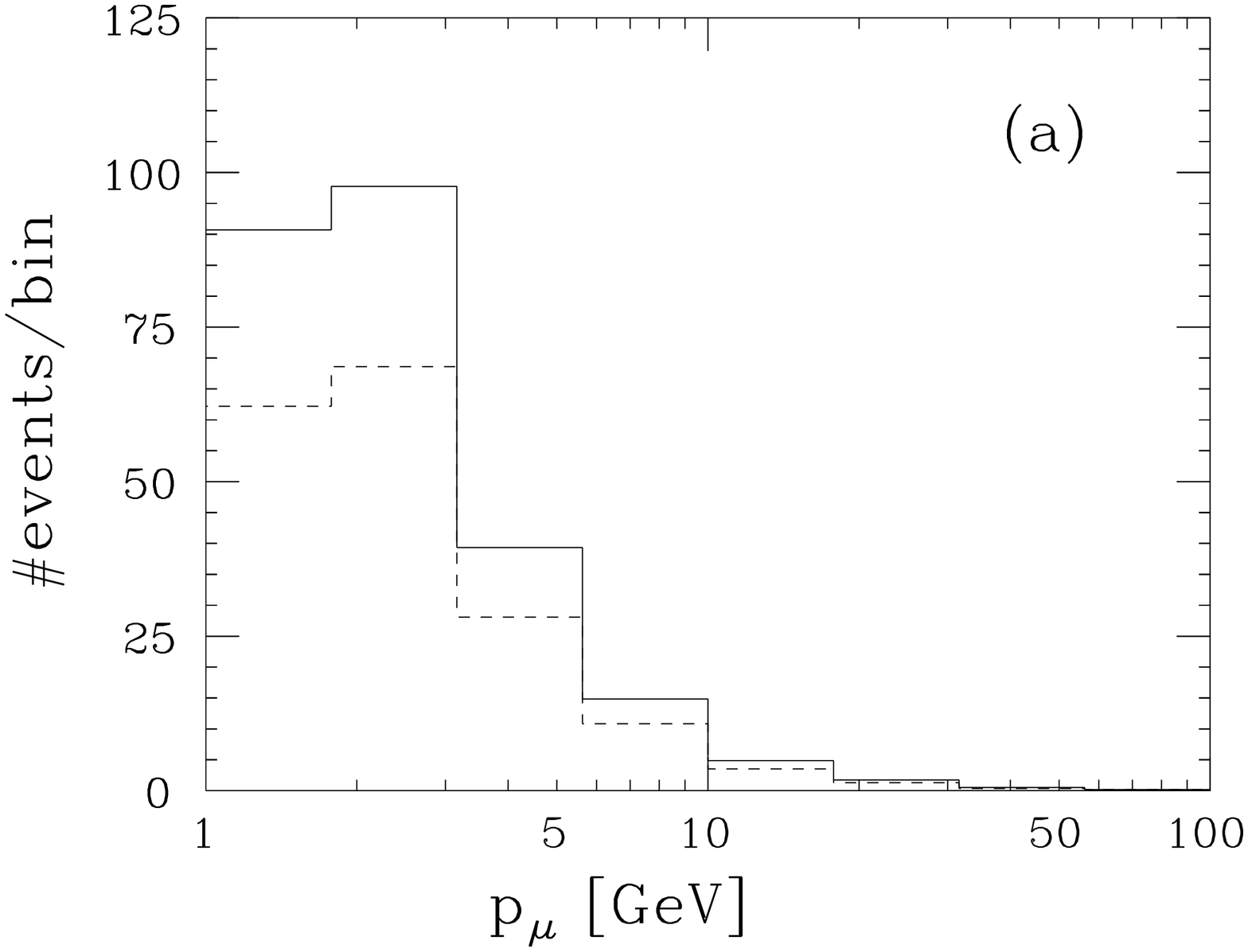,width=0.45\textwidth,angle=90}
    \psfig{file=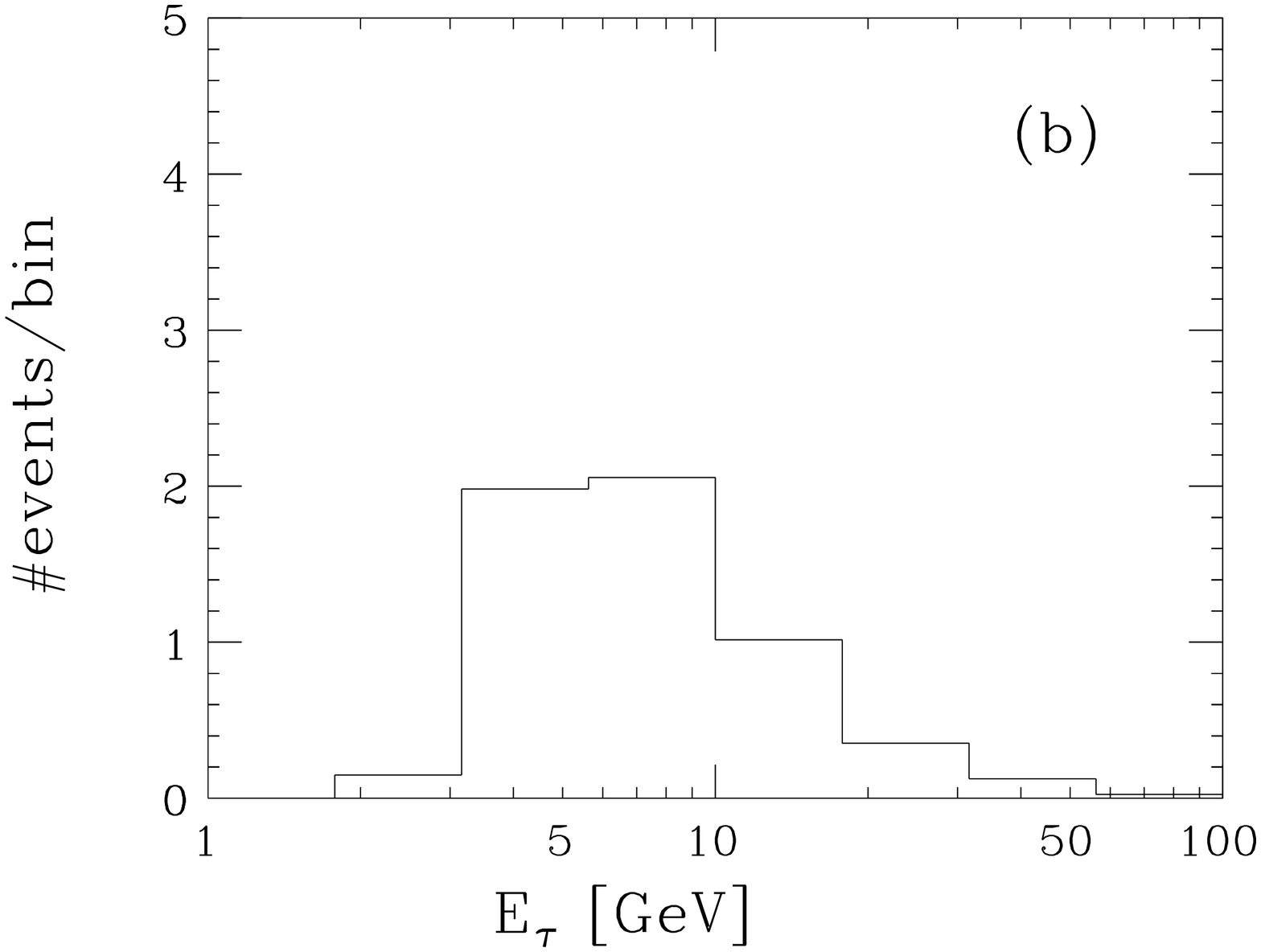,width=0.45\textwidth,angle=90} }
        \caption[tau]{(a) $\mu$ and (b) $\tau$ production rates from
          QE reactions at Super-Kamiokande for 414 days for lepton 
          momentum greater than 1.33~GeV (Multi-GeV).  The dashed
          histogram in (a) and the solid histogram in (b) are the case
          with neutrino oscillation, with parameters given in
          Eq.~(\ref{eq:testpoint}).}
        \label{fig:El}
\end{figure}

We studied the momentum distribution of $e$ and $\mu$ events after
rescaling the normalization as discussed, for a 414 day exposure at
SuperKamiokande (21.1~kton).  This distribution for $\mu$ events is 
shown in Fig.~\ref{fig:El}(a). Compared to the reported data, the 
fraction of QE events in the one-ring event samples becomes smaller 
for higher momentum as expected on physical grounds.  The other events 
correspond to the production of a few additional soft particles below 
the detection threshold.  We will come back to this issue later.

Including $\nu_{\mu}$-$\nu_{\tau}$ oscillations for the ``test point'' 
used by Super-Kamiokande \cite{Kearns-ITP} 
\begin{equation}
        \Delta m^{2} = 0.005~\mbox{eV}^{2}, \sin^{2} \theta = 1,
        \label{eq:testpoint}
\end{equation}
we obtain the dashed histogram in Fig.~\ref{fig:El}(a) 
for the muon momentum distribution.  
Correspondingly, we find $\tau$-appearance from the QE reaction as 
shown in Fig.~\ref{fig:El}(b).  The event rate is 4.43 events for 
414 days.

\begin{figure}
  \centerline{ \psfig{file=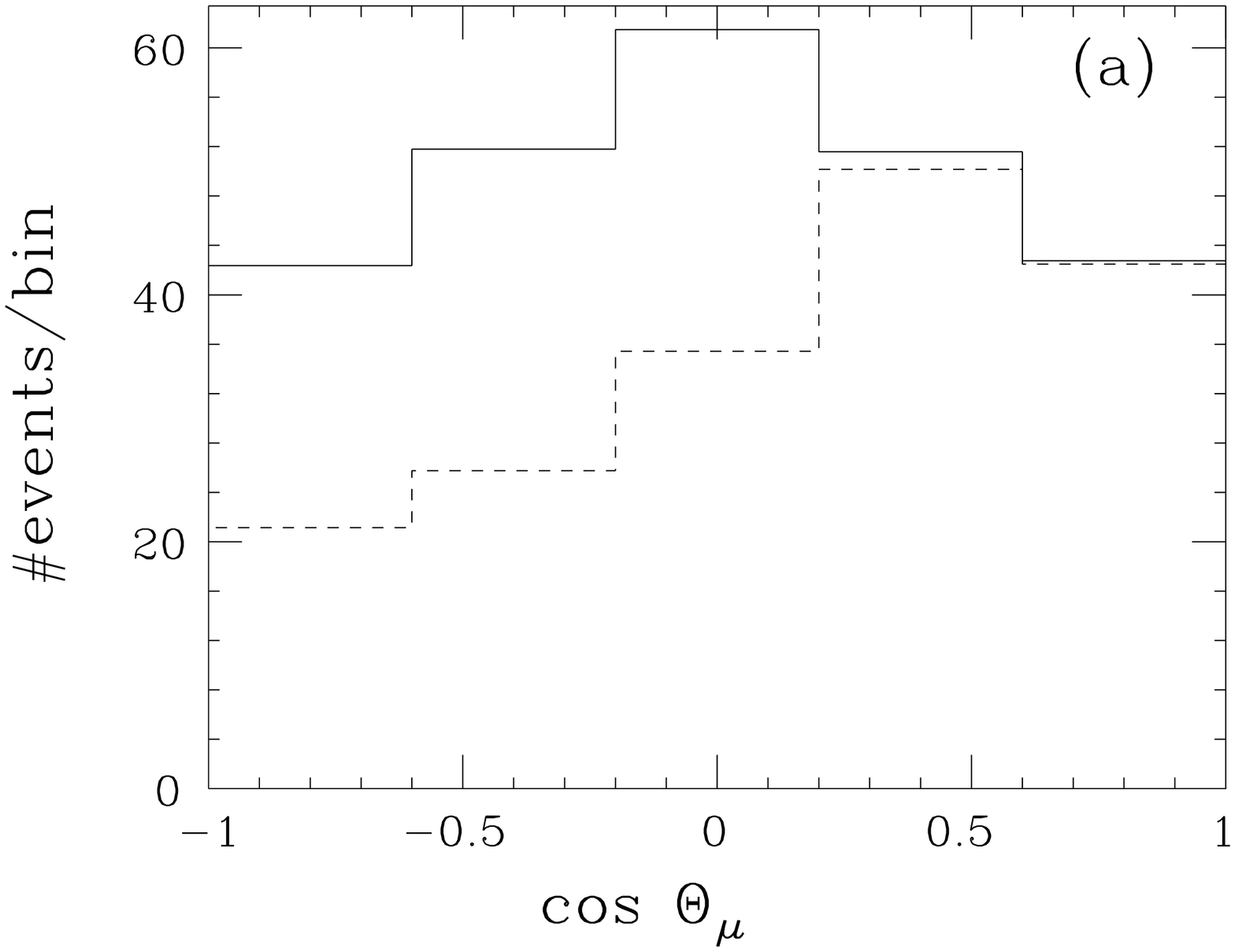,width=0.45\textwidth,angle=90}
    \psfig{file=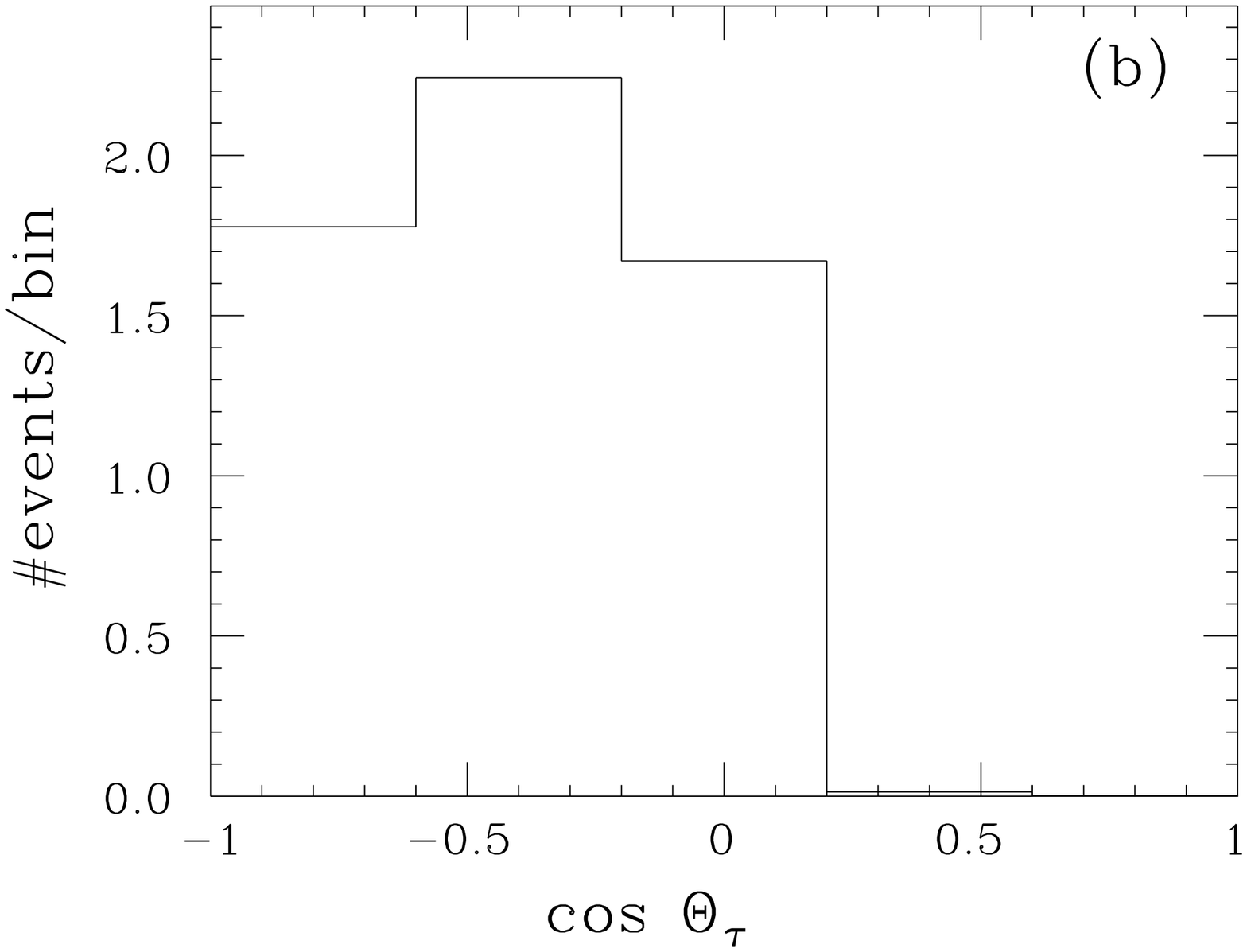,width=0.45\textwidth,angle=90} }
        \caption[tau]{The zenith angle distributions in (a) $\mu$ and 
          (b) $\tau$ production from QE reactions at Super-Kamiokande
          for 414 days in the multi-GeV region.  The dashed histogram in (a)
          and the solid histogram in (b) are the case with neutrino
          oscillation, with parameters given in
          Eq.~(\ref{eq:testpoint}).}
        \label{fig:zenith}
\end{figure}

The zenith angle distributions from the QE reaction 
are shown for $\mu$ (with and 
without oscillation) in Fig.~\ref{fig:zenith}(a)
and for $\tau$ (from oscillation) in Fig.~\ref{fig:zenith}(b).
As expected, all of the $\tau$ produced are up-going.  

We do not discuss the details of DIS production of $\tau$ because, 
as will be discussed later, such events are hard to identify. Nonetheless, we 
note that the event rate for DIS production is comparable to that of QE 
production.  Therefore the current SuperKamiokande data set must 
contain about 10 $\tau$ events for the oscillation parameters 
(\ref{eq:testpoint}).

\section{Backgrounds}

The DIS production of $\tau$ leads to multi-ring events.  
However, the characteristics of these events make it difficult
to identify the $\tau$.  The main difference from DIS CC production of
$e$ and $\mu$ is the {\it absence}\/ of an energetic charged lepton,
which is difficult to establish in multi-ring events, especially for
$e$.  The DIS NC events do not have charged leptons and are even more
similar to DIS $\tau$ events.  Furthermore,
the DIS cross sections for $\tau$ production in the relevant 
range are smaller than those for the corresponding $\mu$ production by 
an order of magnitude. It appears to be difficult to separate 
the DIS $\tau$ events from the DIS $\mu$ events, and we do not
consider DIS $\tau$ production in this letter.

Instead, we focus on the QE production of $\tau$.  Apart from the 
possible ring of the recoil nucleon, the entire event is caused by 
the decay products of the $\tau$.  The purely leptonic decay modes, 
$\tau \rightarrow \nu_{\tau} \nu_{l} l$ for $l=e, \mu$, have branching 
fractions of 17.8\%, and 17.4\%, respectively.  These events would 
appear the same as the QE production of $e$ or $\mu$, and hence cannot 
be used as the signal.  

The single pion decay mode $\tau^{\pm} \rightarrow \nu_{\tau} 
\pi^{\pm}$ has a branching fraction of 11.3\%.  The physics background 
to this signal is from the single pion production from neutral current 
(NC) interaction of the neutrinos with nucleons,
\begin{eqnarray}
         &  & {\mathop{\nu_{l}}\limits^{(-)}} n \rightarrow 
         {\mathop{\nu_{l}}\limits^{(-)}} p \pi^{-},
        \label{eq:NCpim}  \\
         &  & {\mathop{\nu_{l}}\limits^{(-)}} p \rightarrow 
         {\mathop{\nu_{l}}\limits^{(-)}} n \pi^{+},
        \label{eq:NCpip}
\end{eqnarray}
where ${\mathop{\nu_{l}}\limits^{(-)}}$ can be any of the active 
neutrino species.  In the energy range of our interest, the NC 
production of a single pion is dominated by the production of baryon 
resonances with their subsequent decays into a nucleon and a pion.  
Rein and Sehgal (RS) \cite{RS} attempted the calculation of the 
resonance contributions, assuming a simple harmonic oscillator model 
of the baryon resonances to calculate the transition amplitudes.  Even 
though the model itself is somewhat questionable, their calculations 
show reasonable agreement with data within about 20\% for charged 
current (CC) single pion production \cite{SKAT}.  Therefore, we can 
rely on RS calculations on neutral current single pion production 
cross sections at 20\% level.  The cross sections for processes 
(\ref{eq:NCpim},\ref{eq:NCpip}) are almost flat 
above $E_{\nu} \gtrsim 3$~GeV, both of them at 1.0~fb \cite{RS}.  
Unfortunately, this is larger than the expected size of the signal 
cross sections Eqs.~(\ref{eq:QEtau-},\ref{eq:QEtau+}) once multiplied 
with the tau branching fraction to single pion of 11.3\%.  This process 
also suffers from the ``fake'' background of charged current muon 
production where the muon is identified as a charged pion.  This cross 
section is an order of magnitude larger, and is a continuum in muon 
energy with no features.  Even though the SuperKamiokande 
Collaboration is working on the discrimination of charged pions from 
muons, which 
appears feasible based on their beam-based study at KEK \cite{Sobel}, 
identification of $\tau$ in the pion mode appears quite difficult.

The $\tau \rightarrow \nu_{\tau} \rho^{-}$ mode has a much larger 
branching fraction of 25.2\%.  We find this mode much more promising 
than the pion mode.  Unfortunately, the total cross sections of the 
physics background, the NC single-rho production, are neither measured 
nor calculated.  We therefore extend the RS calculations to rho 
production by summing over the resonance production cross sections 
given in their TABLE IV multiplied by the branching fractions into rho 
from Particle Data Group \cite{PDG96}.  Our calculations neglect the 
interference between resonances as well as the continuum contribution, 
but we have verified that our naive method gives cross sections for 
the single pion production which agrees with their full calculations 
within about 10\%.  The larger uncertainty, however, results from the 
lack of data on the rho branching fractions of baryon resonances.  We 
therefore give conservative and optimistic estimates of the NC 
single-rho production cross sections by taking the upper (up to 100\% 
if not measured) and lower (down to 0\% if not measured) ends of the 
rho branching fractions, respectively.  
The results of our calculations are shown in 
the Table~\ref{tab:rho}.  The uncertainties are unfortunately rather 
large.  It is encouraging, however, that these cross sections are up 
to an order of magnitude lower than the signal cross sections of QE 
$\tau$ production.

\begin{table}[t]
        \caption[rho]{The single rho production cross sections in fb using 
        the resonance production cross sections in \cite{RS} and their 
        subsequent decays into rho in \cite{PDG96}.}\label{tab:rho}
        \centerline{
        \begin{tabular}{|c|cc|}
                \hline
                 & $E_\nu = 2$~GeV & $E_\nu = 20$~GeV \\ \hline
                $\nu p \rightarrow \nu n \rho^+$ & 0.11 -- 0.13 & 0.20 -- 0.50\\
                $\nu n \rightarrow \nu p \rho^-$ & 0.074 -- 0.14 & 0.26 -- 0.57\\ 
                \hline
        \end{tabular}
        }
\end{table}

There are other ``fake'' backgrounds to the single rho production from 
$\tau$ appearance.  The most important one is the charged current 
$\nu_\mu$ event with a single neutral pion production, {\it i.e.}\/ 
$\nu_\mu n \rightarrow \mu^- p \pi^0$ or $\bar{\nu}_\mu p \rightarrow 
\mu^+ n \pi^0$, where the muon is misidentified as a charged pion and 
the invariant mass of the charged pion and the neutral pion falls into 
the range of the rho mass.  The relevant cross sections are 
\begin{eqnarray}
        \sigma (\nu_{\mu} n \rightarrow \mu^{-} p \pi^{0}) & \simeq & 
        3~\mbox{fb},
          \\
        \sigma (\bar{\nu}_{\mu} p \rightarrow \mu^{+} n \pi^{0}) & \simeq & 
        2~\mbox{fb} .
\end{eqnarray}
Comparing with the signal cross sections, 
Eqs.~(\ref{eq:QEtau-},\ref{eq:QEtau+}), and $\tau \rightarrow 
\nu_{\tau}\rho$ branching fraction of 25.2\%, we would like to 
achieve a rejection factor on $\mu^{\pm} \pi^{0}$ production by an 
order of magnitude (keeping $S/N \gtrsim 3$).  Even though we do not 
know the realistic number for it, it is reasonable to expect that the 
combination of the mass cut on $\pi^{\pm} \pi^{0}$ within the 
$m_{\rho}$ range together with $\pi^{\pm}/\mu^{\pm}$ discrimination 
being discussed \cite{Sobel}, this level of rejection can be 
achieved.

The next class of background events to discuss is the deep inelastic 
events with $\rho$ production.  The inclusive production rate of 
$\rho^{\pm}$ from neutral current reactions is not measured.  
However, there is a reported rate of inclusive production rate of 
$\rho^{0}$ from charged current reactions, which give $\rho$ 
multiplicity of about 0.1 \cite{SKAT2} in the energy range of our interest.  
Assuming the comparable $\rho^{\pm}$ production rate from neutral 
current reactions, together with the total neutral current cross 
section of about 7~fb at $E_{\nu} \sim 5$~GeV, the inclusive $\rho$ 
production cross section would be about 0.7~fb.  Since these events 
are likely to be associated with production of a few other pions, the 
requirement of {\it single}\/ $\rho$ production is expected to 
suppress this background further.  Overall, the background rate from 
deep inelastic $\rho$ production is much smaller than the signal.

Another possible source of background is the neutral current events of
multi-pion production (DIS) where only one neutral pion is produced
and only one charged pion is above the \v{C}erenkov threshold.  The
estimate of this background depends sensitively to the detector
capability and is beyond the scope of this paper.  It should be noted,
however, that there is no particular reason for the neutral and
charged pions to form an invariant mass in the rho mass range for
this process, and the continuum DIS process itself is relatively low
at this energy range.

As we have seen above, the estimates of the background are quite crude 
at best, primarily because of the lack of relevant data at this energy 
region.  

The efficiency in reconstructing rho is crucial in the study of $\tau$ 
appearance.  The closest study we could find in the literature is the 
search for proton decay $p \rightarrow \rho^+ \nu$ by Kamiokande 
experiment \cite{Kam-pdecay}.  The efficiency was unfortunately low: 
15\%.  We can only hope that the Super-Kamiokande can achieve higher 
efficiency thanks to much better granularity of the detector.  It is 
encouraging to quote the high efficiency of 54\% quoted by the IMB 
collaboration in their search for $p \rightarrow \rho^{\pm} \nu$ 
\cite{ICHEP-IMB}.  

Recall that we discussed only the QE contribution to $\tau$
production.  There are, however, additional processes of $\tau$
production with additional particles where they are below the
\v{C}erenkov threshold.  This adds more events to the single $\tau$
production.  On the other hand, the same enhancement occurs also for
the background.  It is not clear what the exact situation is.

Let us emphasize that the background cross section can be measured by
the near detector in the K2K experiment, the long-baseline experiment
from KEK to Super-Kamiokande.  The near detector consists of a 1~kt 
water \v{C}erenkov detector followed by a tracking device, a calorimeter
and a muon chamber.  With the water \v{C}erenkov detector they can study
the ``single-$\rho$'' events.  Then the uncertainty in the estimated
background level will be significantly reduced.

\section{Statistical Significance of $\tau$ Appearance}

We estimate the statistical significance for a $\tau$
appearance signal at Super-Kamiokande. Suppose that the actual total
event rate for signal and background $\rho^\pm$ events is $S$ and $B$
events per year, and that a fraction $\epsilon$ of these are
reconstructed. The background events will be up-down symmetric,
whereas the signal events are dominantly up-going. We assume a value for
$\Delta m^2$ such that all signal events are up-going. 
With $N$ years of data there will be $N \epsilon B/2$ down-going
events and  $N \epsilon (B/2+S)$ up-going events. 

For quasi-elastic $\tau$ production we have calculated $S=1.0$ for
$\sin^2 2 \theta = 1$; while for single $\rho$ production we estimate
$B=0.4$, using the extension of the Rein-Sehgal resonance calculation
discussed in the previous section. 
Hence with $\epsilon = 0.5$, and $N=4 (8)$ we expect 0.4 (0.8)
downgoing events and 2.4 (4.8) up-going events. We would like to
determine the probability that the up-going events are caused by a
statistical fluctuation of the background; however, it is difficult to
do this precisely because there are so few down-going events that the
background is very poorly determined. Hence we will assume that the
background has been determined from measurements in a neutrino
beamline, such as at the near detector of the K2K experiment. In
this case, for the observed excess of up-going events compared to
down-going events to be due to a statistical fluctuation of the
background requires a fluctuation of $\sqrt{2N\epsilon} \; S/\sqrt{B}$
standard deviations. With $S=1$ and $B=0.4$, this gives
a 4.6$\sigma$ effect for $N \epsilon = 4$.
As discussed earlier, there are other contributions to both $S$ and
$B$, so that it is probable that the $S/\sqrt{B}$ value used here,
1.6, is not correct. 
For  $S/\sqrt{B} = (1.0, 1.6, 2.5)$, the signal is a $(2.8, 4.6, 7.1) \sigma$
effect, for $N \epsilon = 4$.
Hence a significant result is only expected from a multi-year search
with high $\rho$ reconstruction efficiency. The event rates are low,
so that in practice the signal could
be worse or better depending on whether the number of
observed events happens to fluctuate above or below the mean
expected number.

\section{Conclusion}

We have studied the prospect of detecting $\tau$ appearance in 
Super-Kamiokande detector, from the oscillation of atmospheric 
$\nu_{\mu}$ to $\nu_{\tau}$, and have identified $\tau \rightarrow 
\nu_{\tau} \rho^{\pm}$ as a promising detection mode.  
The event rate, however, is low: about 1.0 event per year from 
quasi-elastic production.  There are possibly some additional 
contributions from non-QE processes where the other products (pions) 
are below the Cerenkov threshold.  We estimated rates for background 
processes, and, even though the estimates have large uncertainties, 
none of the background processes appear larger than the signal.  The 
$S/N$ can range anywhere between $\sim 1/3$ to 1.  This uncertainty 
will be reduced by direct measurements of the background at the K2K 
experiment. Although challenging, with 5 to 10 years of data, the 
detection of $\tau$ appearance from atmospheric neutrinos may be
possible, for example, with an observation of 5 upward going events
and 1 downward.

\section*{Acknowledgements}
HM thanks Michael S. Chanowitz, Stuart J. Freedman, Michael Jones,
Takaaki Kajita, Kam-Biu Luk, Gil Shapiro, and Henry W. Sobel
for useful discussions.  This work was supported in part by the U.S.
Department of Energy under Contracts DE-AC03-76SF00098, in part by the
National Science Foundation under grant PHY-95-14797.  HM was also
supported by Alfred P. Sloan Foundation.

\end{document}